\documentclass[pre,a4paper,superscriptaddress,twocolumn,showpacs,amsmath,amssymb,floatfix]{revtex4}

\usepackage{graphicx}
\usepackage{amssymb}
\usepackage{float}

\usepackage{color}

\def\<{{<}}
\def\>{{>}}

\begin{document}
	
	\title{Jaynes-Cummings model under monochromatic driving}
	
	\author{Leonardo Ermann}
	\affiliation{Departamento de F\'{\i}sica Te\'orica, GIyA,
         Comisi\'on Nacional de Energ\'{\i}a At\'omica.
           Av.~del Libertador 8250, 1429 Buenos Aires, Argentina}
         \affiliation{Consejo Nacional de Investigaciones
             Cient\'ificas y T\'ecnicas (CONICET), Buenos Aires, Argentina}
	\author{Gabriel G. Carlo}
	\affiliation{Consejo Nacional de Investigaciones
             Cient\'ificas y T\'ecnicas (CONICET), Buenos Aires, Argentina}
        \author{Alexei D. Chepelianskii}
        \affiliation{LPS, Universit\'e Paris-Sud, CNRS, UMR 8502, Orsay F-91405, France}
	\author{Dima L. Shepelyansky}
	\affiliation{\mbox{Laboratoire de Physique Th\'eorique, IRSAMC, 
			Universit\'e de Toulouse, CNRS, UPS, 31062 Toulouse, France}}
	
	\date{March 21, 2020}
	
	\begin{abstract}
	We study analytically and numerically 
        the properties of Jaynes-Cummings model under monochromatic driving.
        The analytical results allow to understand the regime of two
        branches of multi-photon excitation in the case of close resonance between
        resonator and driven frequencies. The rotating wave approximation
        allows to reduce the description of original driven model to an effective
        Jaynes-Cummings model with strong coupling between photons and qubit.
        The analytical results are in a good agreement with the numerical ones
        even if there are certain deviations between the theory and numerics
        in the close vicinity of the resonance. We argue that the rich properties of driven
        Jaynes-Cummings model represent a new area for experimental investigations
        with superconducting qubits and other systems.
	\end{abstract}

	%

	\maketitle
	
	\section{Introduction} 
	\label{sec1}
	The Jaynes-Cummings model (JCM) \cite{jc} is the cornerstone system of quantum optics
describing interactions of resonator photons with an atom, 
considered in a two-level approximation.
The usual experimental conditions correspond to a weak coupling constant 
between photons and atom. In this regime the quantum evolution of the system is integrable 
demonstrating revival energy exchange between photons and atom
\cite{jc,eberly,eberlybook,scully}. Such revival behavior had been first observed in experiments
with Rydberg atoms inside a superconducting cavity \cite{walther}.
The overview of applications of JCM for various physical systems is given in 
\cite{jkono,noriphysrep}.

With the appearance of long living superconducting qubits
\cite{esteve} the coupling of such a qubit (or an artificial two-level atom)
to microwave photons of cavity quantum electrodynamics (QED resonator or oscillator) 
 became an active field of
experimental research \cite{wendin}.
Thus single artificial-atom lasing \cite{astafiev}
and a nonlinearity of QED system \cite{fink} have been realized and 
tested  experimentally.
In the frame of QED coupling between qubit and resonator
it is very natural to consider the case of 
resonator pumping by a  monochromatic microwave field
(see e.g. \cite{astafiev,ilichev,buisson}).
Thus the problem of monochromatically driven resonator with photons
coupled to a qubit represents an interesting fundamental extension
of JCM. This system can be viewed as a quantum monochromatically 
driven oscillator coupled to a qubit (or two-level atom or spin-1/2).

The first studies of JCM under monochromatic driving
had been performed for the case of a dissipative 
quantum oscillator studied numerically in the frame of
quantum trajectories \cite{zsprl}.  It was shown that 
under certain conditions the qubit
is synchronized with the phase of monochromatic driving
providing an example of quantum synchronization
in this, on a first glance, rather simple system.
The unusual regime of bistability induced by quantum
tunneling has been reported   
which still requires a better understanding \cite{zsprl,mavrogordatos}.
It was shown that many photons can be excited even at
a relatively weak driving amplitude.
It was also shown that  two different qubits can be synchronized
and entangled by the driving under certain conditions \cite{zsprb}.
Thus the driven JCM represents a very interesting example
of a fundamental problem of quantum 
synchronization \cite{zssync}. From the discovery
of synchronization by  Christian Huygens in 1665 \cite{huygens}
this fundamental nonlinear phenomenon has been 
observed and studied in a variety of real 
systems described by the classical dynamics \cite{pikovsky}. 
At present the development of quantum technologies and especially superconducting qubits
led to a significant growth of interest to the phenomenon
of quantum synchronization 
(see e.g. \cite{bruder1,bruder2,bruder3} and Refs. there in).
Thus the interest to the JCM under driving is growing with 
appearance of new experiments (see e.g. \cite{devoret,paraoanu,eschner}). 
The theoretical investigations by different groups
are also in progress \cite{zsprl,mavrogordatos,fischer,nori}.

We note that the unitary evolution of driven JCM
has been considered in \cite{refereeref1} 
in the rotating wave approximation (RWA)
for the specific resonance case showing that above a certain driving
border the Floquet eigenstates are not normalizable.
In \cite{refereeref2} the comparison was done between the RWA
and non-RWA evolution has been considered showing the existence of
certain difference between these two cases. 

With the aim of deeper understanding of the properties of driven JCM 
we study here the nondissipative case when the system evolution is
described by the quantum time-dependent Hamiltonian
and the related Schrodinger equation.
We present here the comparative analysis of analytical
and numerical treatment of this system.
We develop the semiclassical description of quantum evolution 
considering mainly the case of close (but not exact)
resonance driving with high excitation of oscillator states.

The paper is organized as follows: in Section~\ref{sec2}, we give the system description,
the analytical analysis is described in Section~\ref{sec3},
the numerical results are presented   in Section~\ref{sec4},
the time evolution of coherent states is described in Section~\ref{sec5}, 
discussion of results and conclusion are given in
Section~\ref{sec6}. Appendix provides additional complementary material.
	
\section{System description} 
\label{sec2}

The monochromatically driven JCM is described by the Hamiltonian already considered in \cite{zsprl}:
\begin{equation}
 \hat{H}=\omega_0 \hat{n}+\frac{\Omega}{2} 
 \hat{\sigma}_z+ g\omega_0
(\hat{a}+\hat{a}^\dagger)\hat{\sigma}_x+f\cos{(\omega t)} 
(\hat{a}+\hat{a}^\dagger)
\label{eq1}
\end{equation}
where $\hat{\sigma}_i$ are the usual Pauli operators
describing a qubit,
$g$ is a dimensionless coupling constant,
the driving force amplitude and frequency are 
$f$ and  $\omega$,
the oscillator frequency is $\omega_0$ and $ \Omega$
is the qubit energy spacing. 
The operators $\hat{a},\hat{a}^\dagger$ describe the quantum oscillator
with number of photons being 
$\hat{n}=\hat{a}^\dagger\hat{a}$ ($\hat{n}\vert n\rangle=n\vert n\rangle$).
Here and in the following we take $\hbar=1$.

In the RWA
the Hamiltonian (\ref{eq1}) takes the form:
\begin{equation}\label{eq2}
 \hat{H}=\omega_0 \hat{n}+\frac{\Omega}{2} \hat{\sigma}_z
 + g\omega_0
(\hat{a}\hat{\sigma}_+ +\hat{a}^\dagger\hat{\sigma}_-)+\frac{f}{2}
(\hat{a}e^{i\omega t}+\hat{a}^\dagger e^{-i\omega t}) \; .
\end{equation}

The Floquet theory can be applied to 
the  time periodic Hamiltonians
(\ref{eq1}) and (\ref{eq2}) 
that gives the
Floquet eigenstates ($\vert\Psi_j(t)\rangle$) and Floquet modes 
($\vert\Phi_j(t)\rangle$) 
\begin{equation}\label{eq3}
 \vert\Psi_j(t)\rangle=\exp{(-i\varepsilon_jt/\hbar)}\vert\Phi_j(t)\rangle
\end{equation}
where $\varepsilon_j$ are quasienergy levels defined in the interval $[0,2\pi/T]$ and 
$\vert\Phi_j(t)\rangle=\vert\Phi_j(t+T)\rangle$ are periodic in time.

In the rotating frame the time dependence can be eliminated. 
Thus a state  $\vert\Psi\rangle$, evolving via the Schrodinger equation 
$i\hbar\partial_t \Psi=H \Psi$, can be transformed to 
$\vert\tilde{\Psi}\rangle=
\hat{U}^\dagger\vert\Psi\rangle=\exp\left(i\hat{A}t/\hbar\right) \vert\Psi\rangle$ 
where $\hat{U}^\dagger$ is a unitary operator generated by a Hermitian 
operator $\hat{A}=\omega(\hat{a}^\dagger \hat{a}+\hat{\sigma}_+\hat{\sigma}_-)$. 
Then the system in the rotating frame of RWA
is described by the transformed stationary Hamiltonian
\begin{equation}\label{eq4}
 \hat{H}_{r}=\Delta_0 \hat{n}+\frac{\Delta_\Omega}{2} 
 \hat{\sigma}_z+ g\omega_0
(\hat{a}\hat{\sigma}_+ +\hat{a}^\dagger\hat{\sigma}_-)+
\frac{f}{2}(\hat{a}+\hat{a}^\dagger)
\end{equation}
with $\Delta_0=\omega_0-\omega$ and $\Delta_\Omega=\Omega-\omega$. 
In the following we mainly discuss a 
typical set of  system parameters being $\omega_0=1$, $\Omega=1.2$, $g=0.04$ 
and $f=\lambda\sqrt{n_p} = 0.02 \sqrt{20} =5^{-\frac{3}{2}}\simeq0.0894$ 
(this corresponds to the main set of parameters
 $\lambda=0.02$ and $n_p=20$ discussed in \cite{zsprl} for the dissipative 
case with the dissipative constant $\lambda$ for oscillator).
We check also other parameter sets ensuring that the main set 
corresponds to a typical situation. Below in our studies we use dimensionless units
for parameters being proportional to frequencies ($\omega_0$, $\omega$, $\lambda$ given in Figs.),
the physical quantities are restored from the ratios $ \omega/\omega_0 , \lambda/\omega_0$,
the physical values of system energies are obtain by 
multiplication of reported energies by $ \omega_0 \hbar$
where $\hbar$ is the Planck constant,

\begin{figure}[t]
\begin{center}
\includegraphics[width=0.46\textwidth]{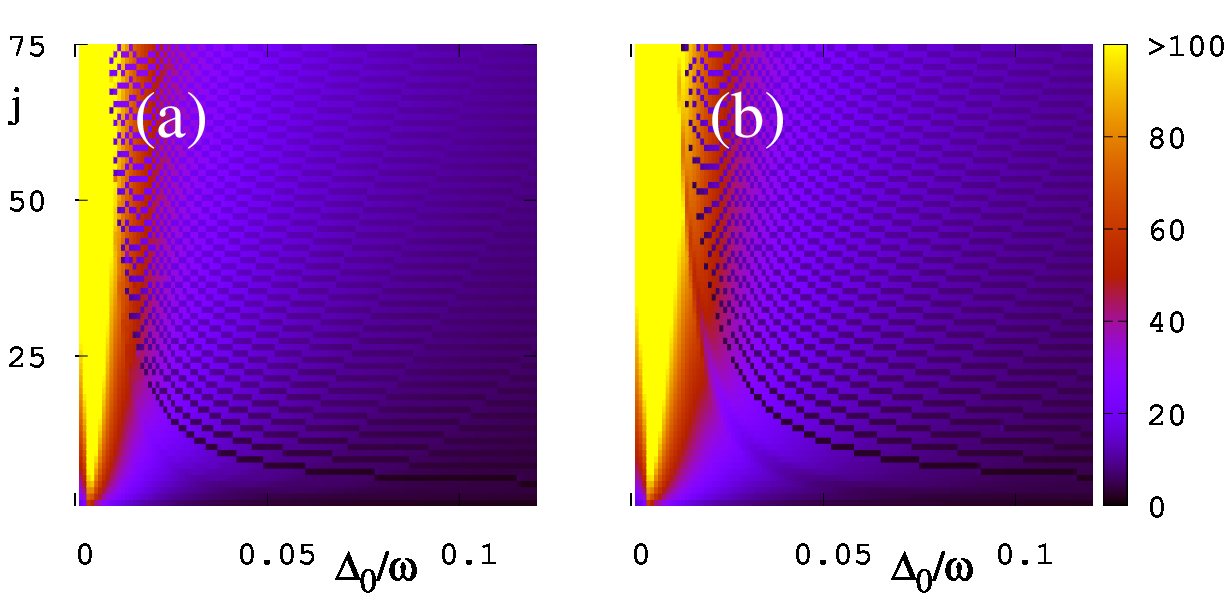}
\end{center}
\vglue -0.3cm
\caption{\label{fig1}
Participation ratio $\xi$ 
of eigenstates $\psi_j$ of RWA Hamiltonian (\ref{eq4}) 
as a function of rescaled resonance detuning $\Delta_0/\omega$ and 
eigenstate index $j$ which counts eigenenergies in their
monotonically increasing order; here $f=5^{-\frac{3}{2}}\simeq0.0894$,
$g=0.04$ and $g=0.08$ in 
left $(a)$ and right $(b)$ panels respectively;
$xi$ values are shown by color with the corresponding color bar.
Here and in all other Figures we use dimensionless units
explained in the text.
}
\end{figure}

The eigenstates $\psi_j$ of RWA Hamiltonian (\ref{eq4}) are determined by the equation
$\hat{H}_{r} \psi_j (n,\sigma_z) = E_j \psi_j(n,\sigma_z) $. We order the index $j$ 
in such a way that the energy eigenvalues $E_j$ are monotonically growing
with $j$.
	
The numerical computation of eigenstates $\psi_j$ is done by a direct matrix diagonalization
with a truncated basis of oscillator eigenstates with $0 \leq n \leq N-1$.
We checked that the value of $N=700$ is sufficient to have stable eigenstates
with $j<100$ so thus the following  numerical results are obtained with this $N$ value.
Thus, with qubit,  in total we have $2N=1400$ states.
We also use the same $N$ to obtain the time evolution of initial Hamiltonian (\ref{eq1}).
The time evolution is obtained by the Trotter decomposition 
with the time step $\Delta =0.005$ (the results are not sensitive to 
further decrease of the time step). 

\begin{figure}[t]
\begin{center}
\includegraphics[width=0.46\textwidth]{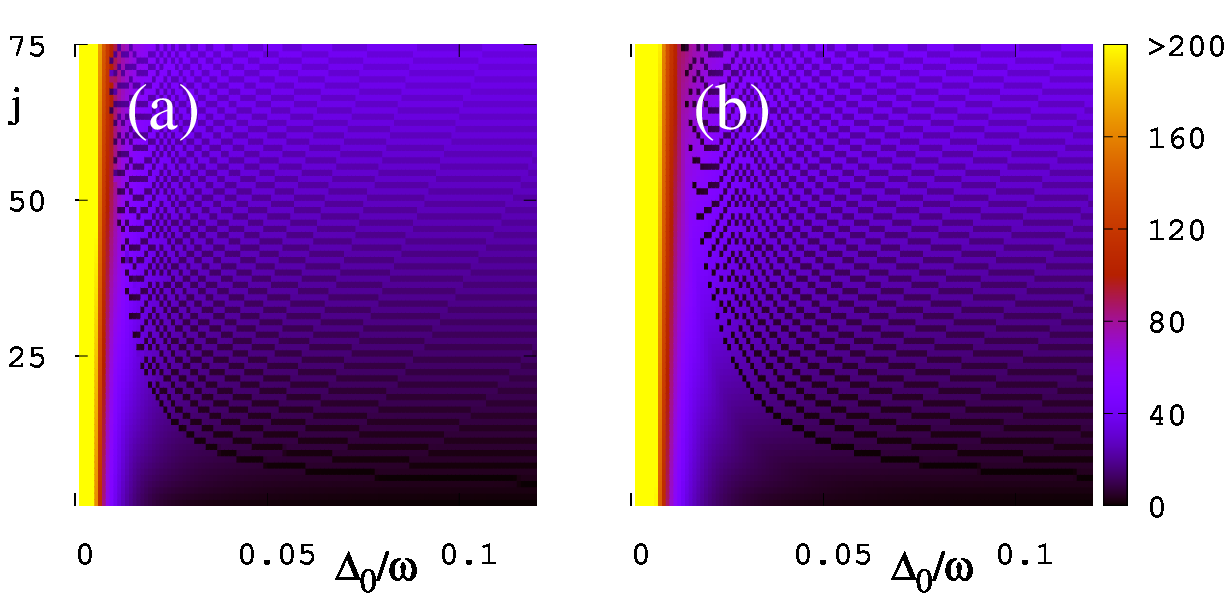}
\end{center}
\vglue -0.3cm
\caption{\label{fig2}
Average oscillator number $<n>$ for eigenstates of Hamiltonian (\ref{eq4})
shown by color for the parameters of Fig.~\ref{fig1} with $g=0.04$ and $g=0.08$ 
in left $(a)$ and right $(b)$ panels respectively. 
}
\end{figure}

We characterize the eigenstates of $H$ (\ref{eq1}) and $H_r$ (\ref{eq4})
by their  participation ratio (PR) defined as
$\xi_j = \sum_{n\sigma_z} \vert \psi_j (n,\sigma_z) \vert^2 / 
\sum_{n\sigma_z} \vert \psi_j (n,\sigma_z) \vert^4$.
Here $ \psi_j(n,\sigma_z)$ represents the eigenfunction expansion in the
eigenbasis at  $g=0$. Thus $\xi_j$
gives an effective number of decoupled states (at $g=0$)
contributing to a given eigenstate at $g>0$. 
For a given eigenstate we also compute
the average photon number  $\langle \psi_j\vert \hat{n} \vert \psi_j\rangle = 
\langle n \rangle$ 
and the average qubit (spin) polarization  $<\sigma_z>$.

\begin{figure}
\begin{center}
\includegraphics[width=0.46\textwidth]{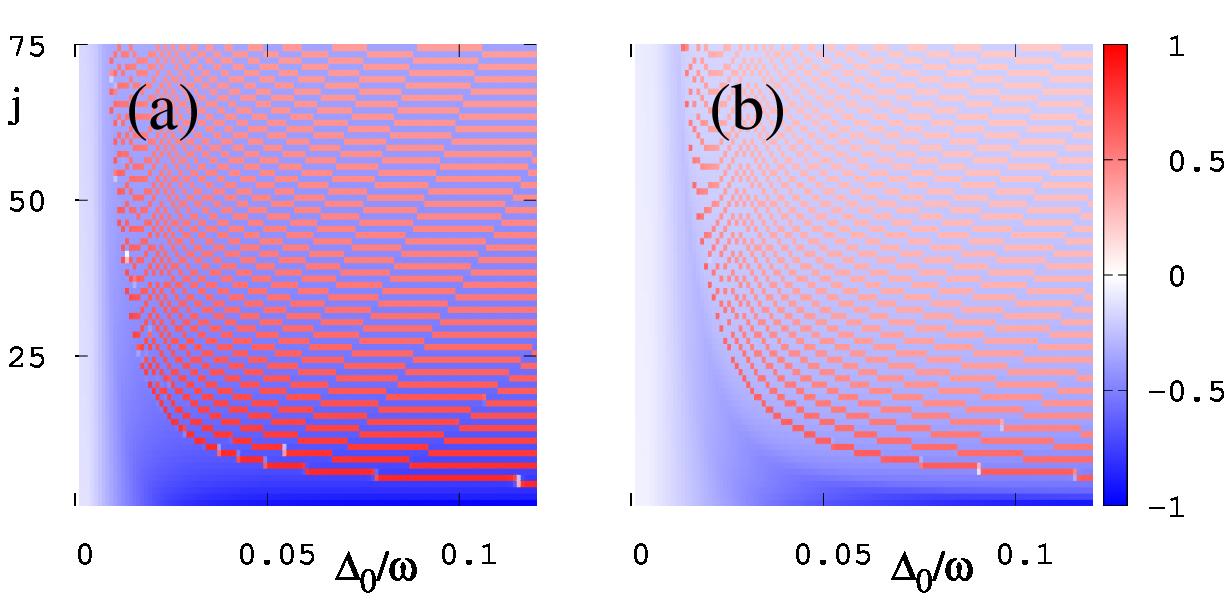}
\end{center}
\vglue -0.3cm
\caption{Average spin $<\sigma_z>$ for eigenstates of Hamiltonian (\ref{eq4})
shown by color for the parameters of Fig.~\ref{fig1} with $g=0.04$ and $g=0.08$ 
in left $(a)$ and right $(b)$ panels respectively. 
}
\label{fig3}
\end{figure}

The dependencies of $\xi$, $<n>$, $<\sigma_z>$, 
for eigenstates $\psi_j$ of Hamiltonian (\ref{eq4}),
on $j$ and rescaled detuning frequency $\Delta_0/\omega$ are shown in 
Figs.~\ref{fig1},~\ref{fig2},~\ref{fig3} respectively.
These results show that in a vicinity of resonance many oscillator states are populated
that is rather natural. The polarization dependence is 
more tricky being close to zero in direct resonance vicinity
and becoming mainly negative with detuning increase
and later followed by a polarization change from positive to negative.
We will return to the discussion of these properties in next Sections.

According to the analytical result obtained in the RWA frame in \cite{refereeref1} 
for the case of exact resonance the Floquet eigenstates become fully 
delocalized over all oscillator states 
(being non-normalizable) for $f \ge g$. Our numerical results
confirm this delocalization  both for non-RWA case of Hamiltonian (\ref{eq1})
and for RWA case of Hamiltonian (\ref{eq4}).
These results are presented in  Appendix Fig.~\ref{figA1}.

\section{Analytical results} 
\label{sec3}

For analytical analysis of driven JCM we perform in (\ref{eq4}) an additional transformation 
using the replacement $\hat{a}=\hat{b}-\frac{f}{2\Delta_0}$ that gives us a transformed 
Hamiltonian 
\begin{equation}
\label{eq5}
 \hat{H_{rt}}=\Delta_0 \hat{n}_b+\frac{\Delta_\Omega}{2} \hat{\sigma}_z+ g\omega_0
\left(\hat{b}\hat{\sigma}_+ +\hat{b}^\dagger\hat{\sigma}_-\right)+B_x \hat{\sigma}_x+K \; .
\end{equation}
This shows an appearance of 
an effective field $B_x={fg\omega_0}/({2\Delta_0})$ and a constant term  
 $K= {f^2}/({4 \Delta_0})$. The interesting feature of the expression (\ref{eq5})
is that even for small $g$ values we obtain an effective JCM
with a strong effective values of effective coupling constant
$g_{eff}= g \omega_0/\Delta_0 \gg g$ at  
small resonance detunings $\Delta_0 \ll \omega_0$.

It is important to note that in (\ref{eq5}) we effectively obtain
the JCM with a strong  coupling $g_{eff} > 1$ between oscillator and spin.
In fact it is known that without RWA the original JCM at strong coupling
is characterized by a chaotic dynamics for the corresponding 
classical equations of motion \cite{zaslavsky,milonni}.
In the quantum case such a chaotic dynamics leads to
quantum chaos for several spins interacting with a resonator
with the level spacing statistics as for random matrix theory \cite{graham}.
Thus the monochromatically driven JCM can be used for investigations
of many-spin quantum chaos induced by an effective strong coupling to a resonator.

On the other hand, the semiclassical version of Eq.(\ref{eq4}) 
can be written in spin 1/2 basis as
\begin{eqnarray}
\label{eq6}
H_{sc}&=&\frac{p^2}{2}+\frac{\Delta_0^2x^2}{2}+f\sqrt{\frac{\Delta_0}{2}}x+\\ &+&\left(
\begin{array}{cc}
\frac{1}{2}\Delta_\Omega & \sqrt{\frac{  \Delta_0g^2\omega_0^2}{2}}
\left(x+\frac{i p}{\Delta_0} \right)
 \\ 
\nonumber
\sqrt{\frac{  \Delta_0g^2\omega_0^2}{2}}
\left(x-\frac{i p}{\Delta_0} \right) & -\frac{1}{2}\Delta_\Omega\\
\end{array}
\right) 
\end{eqnarray}
which can be diagonalized, with the corresponding solution:
\begin{eqnarray} 
\label{eq7}
h&=&h_0+f\sqrt{\frac{\Delta_0}{2}}x
\pm\sqrt{\frac{ g^2 \omega_0^2}{\Delta_0}h_0+
\frac{\Delta_\Omega^2}{4}}
\\
\nonumber
h_0&=&\frac{p^2}{2}+\frac{\Delta_0^2x^2}{2} \; .
\end{eqnarray} 
Here $(x.p)$ are 
classical coordinate and momentum of oscillator which mass 
is taken to be unity $m=1$. The linear term in $x$ in (\ref{eq7}) simply
gives a shift of oscillator center position.

\begin{figure}
\begin{center}
\includegraphics[width=0.46\textwidth]{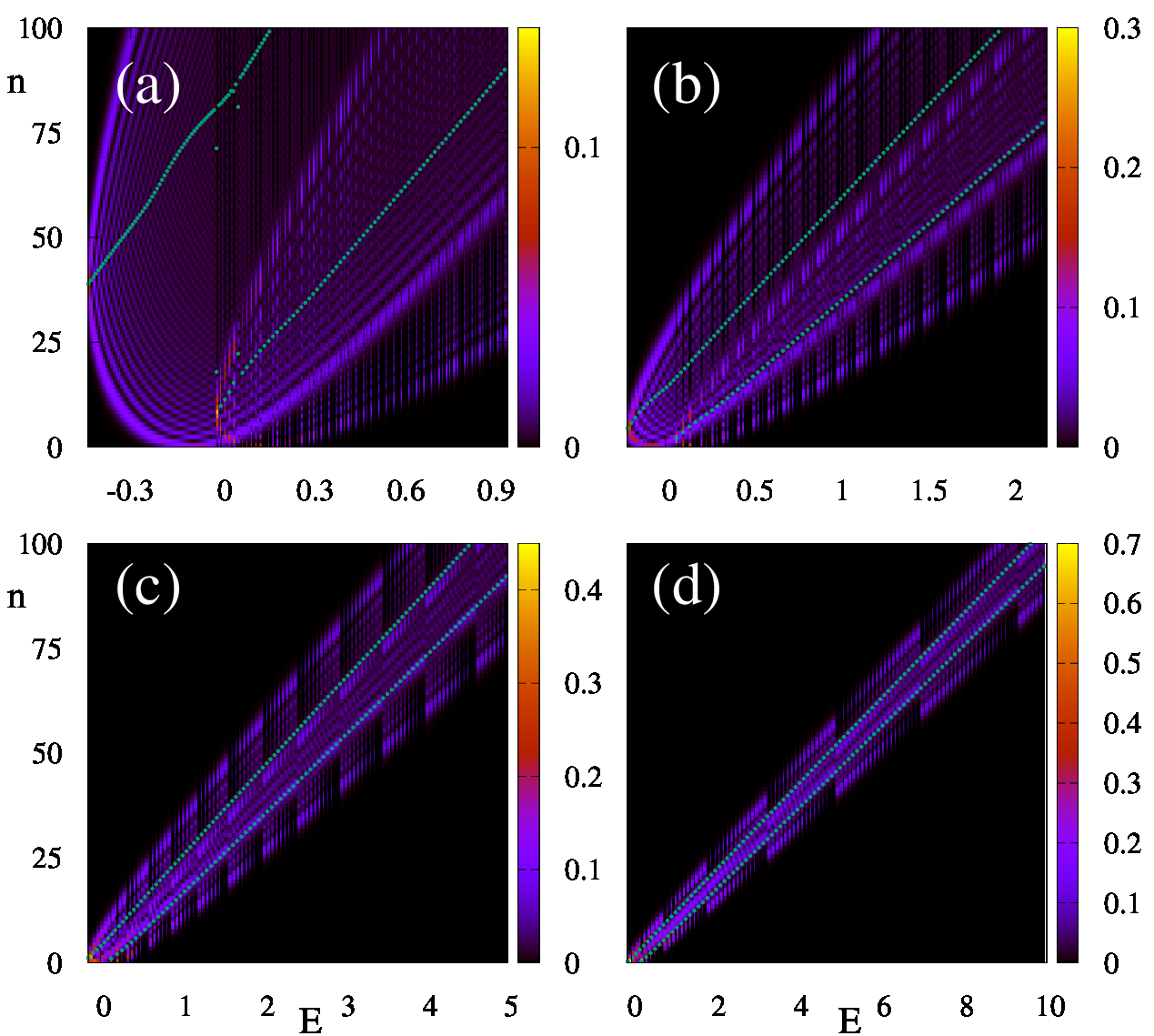}
\end{center}
\vglue -0.3cm
\caption{Probability distribution $P(n)=\vert\langle n\vert\psi_j\rangle\vert^2$ 
(tracing out spin space) of $j^{th}$ eigenstate of (\ref{eq4})
ordered by increasing energy $H\vert\psi_j\rangle=E_j\vert\psi_j\rangle$.
The values of parameters are $g=0.04$, and 
$\Delta_0=\omega_0-\omega=0.01,0.025,0.05,0.1$ 
in panels (a), (b), (c) and (d) respectively.
The green dotted curves show the mean value 
$\langle n\rangle$ of the corresponding
eigenstate $\psi_j$. The color map goes from black at 0
to yellow at maximum value given by $0.14$ for $(a)$, $0.3$ for $(b)$, 
$0.45$ for $(c)$ and $0.7$ for $(d)$. 
}
\label{fig4}
\end{figure}

\begin{figure}
\begin{center}
\includegraphics[width=0.46\textwidth]{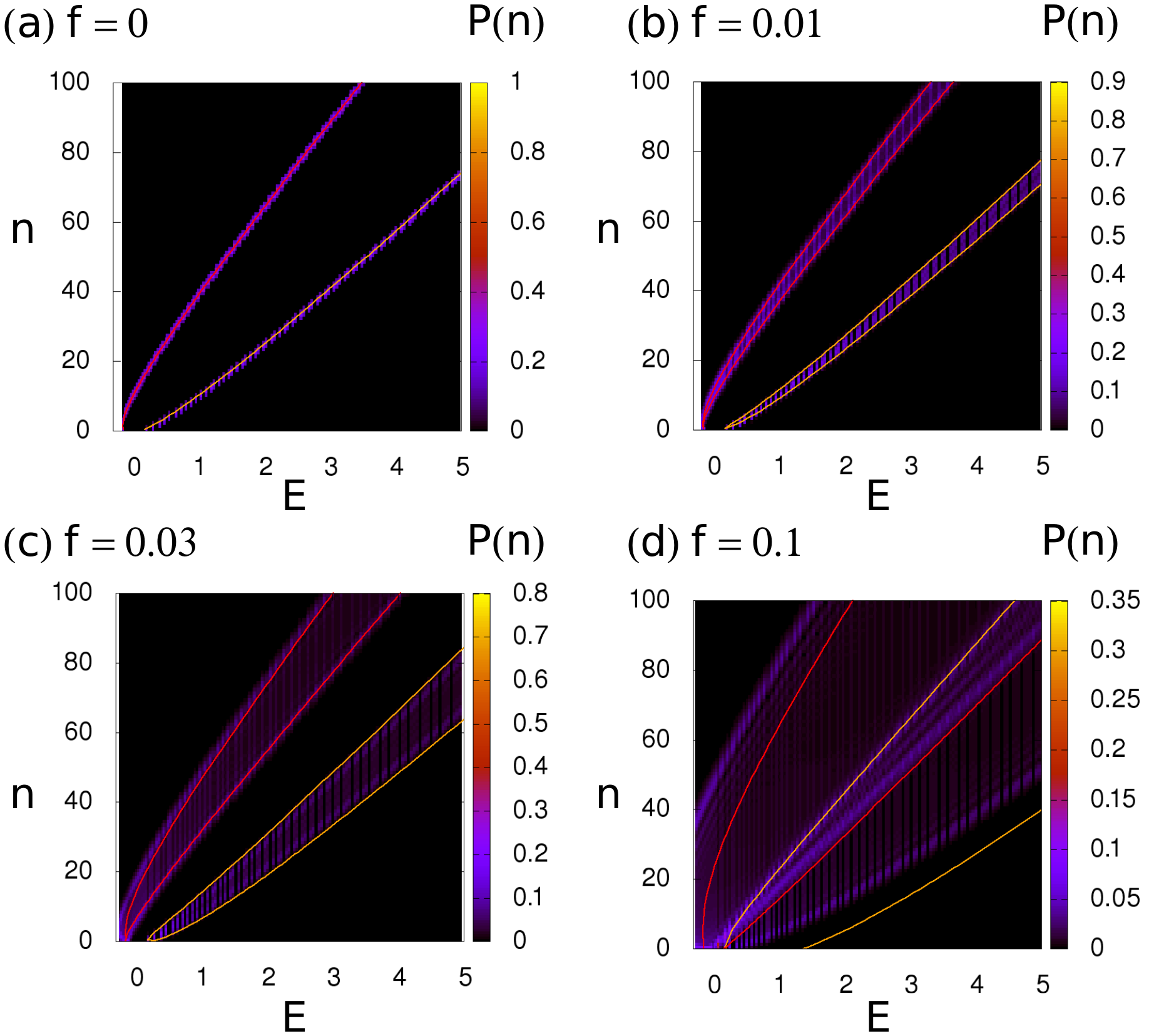}
\end{center}
\vglue -0.3cm
\caption{Probability distribution $P(n)$ of the eigenstates of $\psi$
Eq.~(\ref{eq4}) ordered by increasing energy ${\hat H} \psi = E \psi$ for
different values of the driving force $f$. The other parameters of the
Hamiltonian are set to $\Delta_\Omega = 0.25$, $\Delta_0 = 0.05$, $g
\omega_0 = 0.15$. The red/orange lines show the semi-classical
estimate expressions for the positions of $P(n)$ maxima for the two-spin
eigenstates of the semi-classical hamiltonian given by
Eqs.~(\ref{eqnvse},\ref{eqminmax}). The semi-classical curves are
in a good agreement with the data from quantum wavefunctions and give a physical
interpretation for the position of the maxima of $P(n)$.
}
\label{fig5}
\end{figure}

The above expressions also allow to obtain the semiclassical expression
for the average spin polarization being 
\begin{equation}
\label{eq8}
 \langle\sigma_z\rangle=\pm\left(1+4g^2\omega_0^2\langle n\rangle/\Delta_\Omega^2\right)^{-\frac{1}{2}} .
\end{equation}

The semiclassical theoretical expressions (\ref{eq7})
gives us the dependence of RWA energy $h$ on unperturbed energy $h_0$ which
we compare with the results of numerical simulations in the next Section.
We also compare the theoretical spin polarization (\ref{eq8})  with 
the numerical results.

\section{Numerical results} 
\label{sec4}

The eigenstates of Hamiltonian (\ref{eq4}) are obtained by a direct numerical 
matrix diagonalization with the numerical parameter described above.
The eigenstate probability distribution of $\psi_j$ is shown in Fig.~\ref{fig4} as a function
of oscillator number $n$ and eigenenergy $E=E_j$.
We clearly see the presence of two branches corresponding to two spin polarization.
The mean values of $\langle n\rangle$ are shown by green dotted curves
marking the average dependence $n(E)$ for each branch.
In Appendix Fig.~\ref{figA2}, 
for comparison 
we show the same characteristics as in Fig.~\ref{fig4}
but for eigenstates of transformed Hamiltonian (\ref{eq5}).
We obtain a good agreement between the eigenstates
of these two Hamiltonian confirming the validity of
the analytical transformation from one to another.
At the same time at very small resonance detunings $\Delta_0 = 0.01$
there are certain differences between these two representations
which we attribute to high order corrections in a resonance vicinity.

\begin{figure}
\begin{center}
\includegraphics[width=0.46\textwidth]{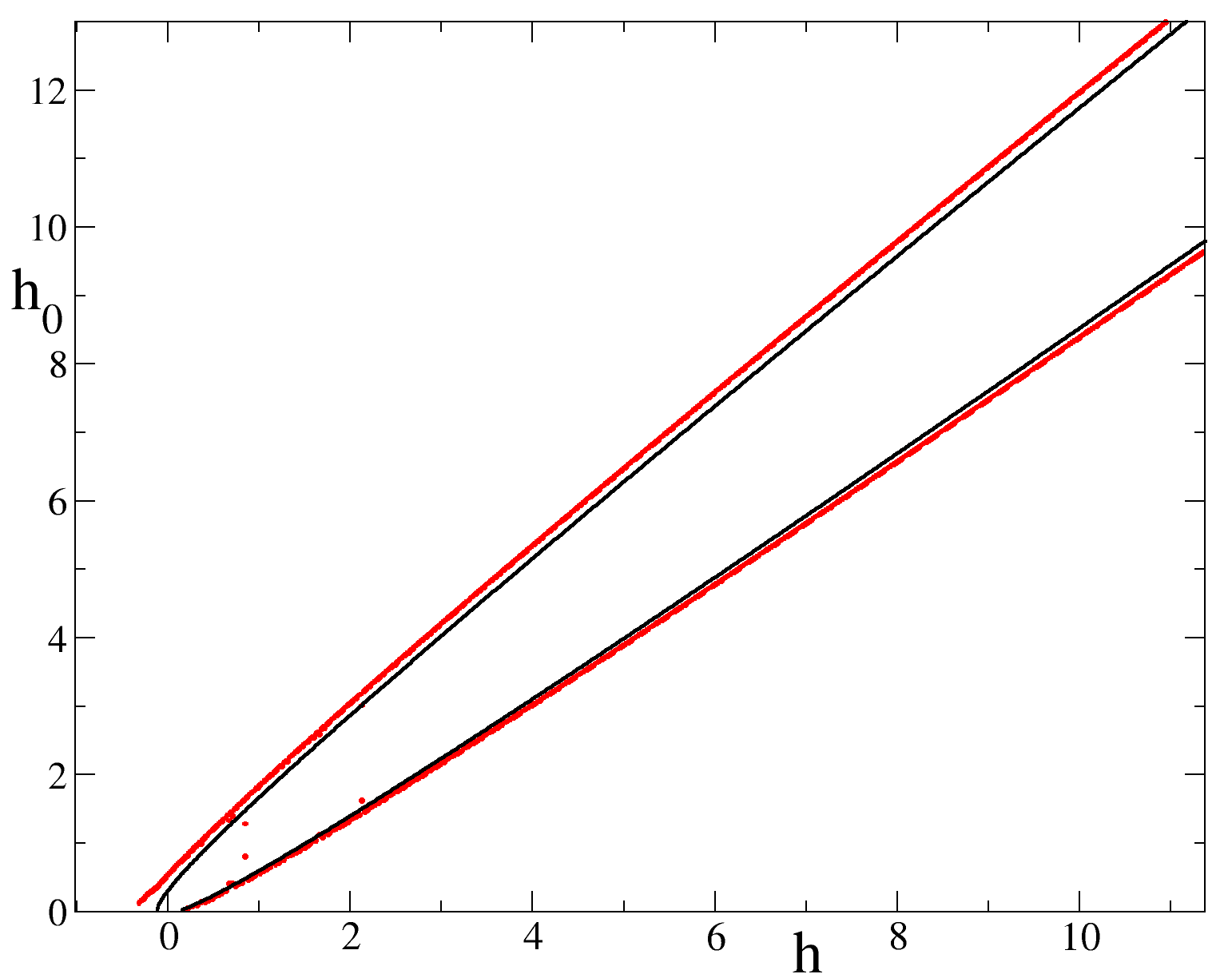}
\end{center}
\vglue -0.3cm
\caption{Quantum harmonic oscillator Hamiltonian $h_0$ vs. $h$ from (\ref{eq7}).
Two branches of (\ref{eq7}) for $f=0$ are shown in black line, while mean values
of $\langle h_0\rangle$ as a function of $\langle h\rangle$ obtained from
the numerical diagonalization of Hamiltonian (\ref{eq4}) are shown by red (gray) dots.
The parameter values are $g=0.04$, $\omega=1$, $\omega_0=0.975$, $\Omega=1.2$, $g=0.04$ 
and $f=\hbar\lambda\sqrt{n_p}$ with $\lambda=0.02$ and $n_p=20$ which are the same as in 
panel (b) of Fig~\ref{fig4}.
}
\label{fig6}
\end{figure}

Averaging the semi-classical Hamiltonian over an oscillation period,
we find the mean oscillator quantum number $\langle n\rangle = h_0 /
\Delta_0$ as function of the eigenstate energy $E = h$ as the positive
solutions of the equation:
\begin{align}
E &= \Delta_0 \langle n\rangle \pm \sqrt{ g^2 \omega_0^2 \langle
n\rangle + \frac{\Delta_\Omega^2}{4}}
\label{eqnvse}
\end{align}
the two possible signs correspond to the two spin-eigenstates of
Eq.~(\ref{eq6}). As can be seen from Fig.~(\ref{fig4}), the probability $P(n)$ is in
general not peaked at its average value  $\langle n\rangle$, instead
(for a fixed eigenstate) $P(n)$ is non zero for n in a range
$(n_{min}, n_{max})$ with maxima at both $n_{min}$ and $n_{max}$. The
position of the maxima can be understood from the following argument.
For simplicity we neglect the change in the Zeeman-like energy term of
Eq.~(\ref{eq6}), in the conservation of energy for semiclassical motion is
then $h \simeq h_0 + f \sqrt{\frac{\Delta_0}{2}} x(t)$ where the
position $x(t)$ follows an oscillation with amplitude $\sqrt{2 \langle
n\rangle/\Delta_0}$. The change in potential energy 
$f \sqrt{\frac{\Delta_0}{2}} x(t)$ is approximately compensated by a
change of $h_0 = \Delta_0 n$. The most likely value of $h_0$
correspond to inflection points of the trajectory giving the estimate:
\begin{align}
n_{max,min} &= \langle n\rangle \pm \frac{f \sqrt{\langle n\rangle}}{\Delta_0}
\label{eqminmax}
\end{align}
This estimation is compared with numerical data on Fig.~(\ref{fig5}) for
different strengths of the force $f$ showing a good agreement with the
roating-wave Hamiltonian wavefunctions. It is interesting that these
simple semi-classical arguments allow to understand some nontrivial
wavefunctions properties of the driven Jaynes-Cummings model
wavefunctions.

The comparison between the numerical results obtained from the eigenstates of Hamiltonian
(\ref{eq4}) and the semiclassical theory of (\ref{eq7}) is also shown in Fig.~\ref{fig6}.
It shows a good agreement between the theory and numerical results.

\begin{figure}
\begin{center}
\includegraphics[width=0.46\textwidth]{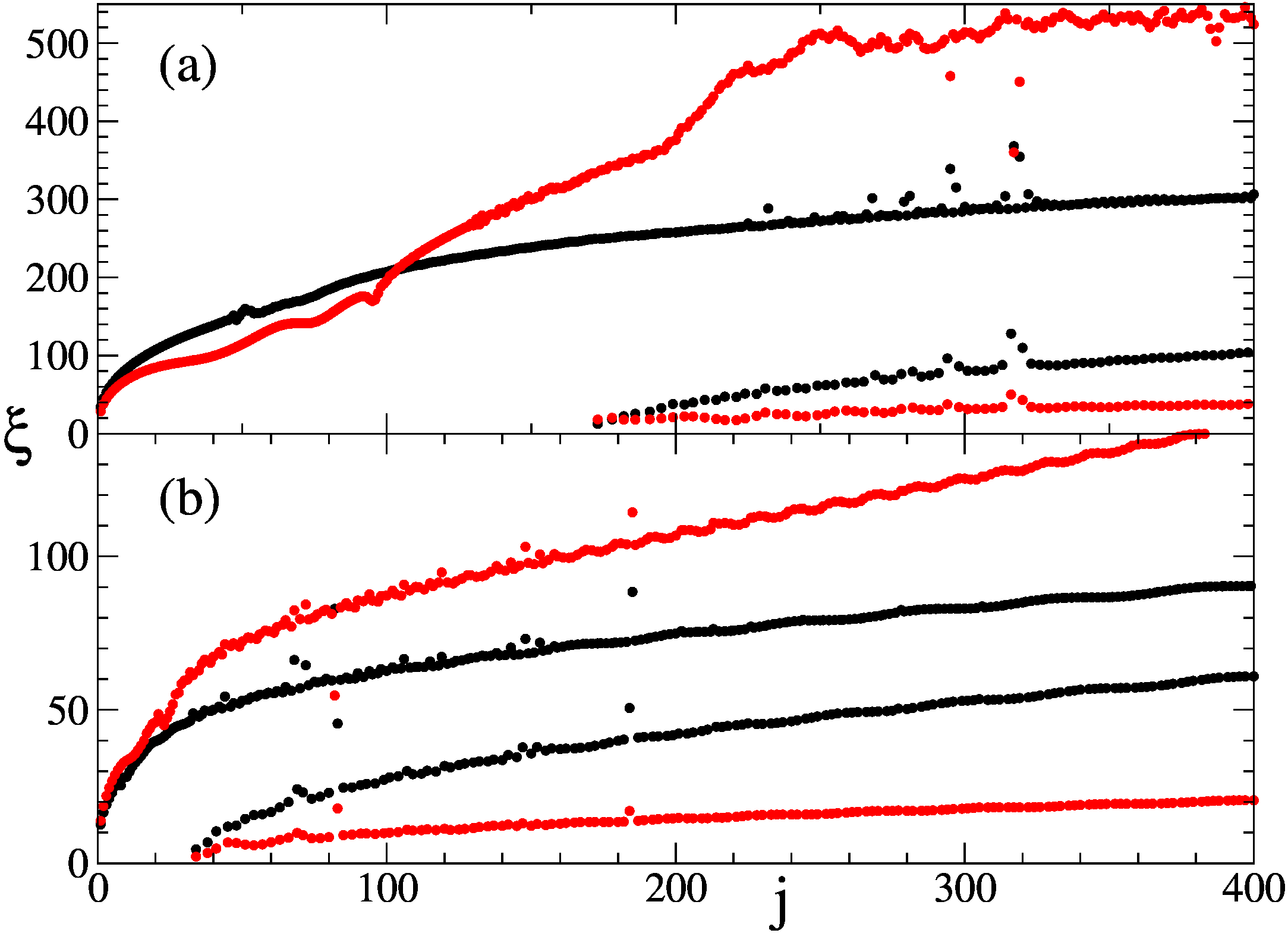}
\end{center}
\vglue -0.3cm
\caption{Participation ratio $\xi$ of eigenstates of Hamiltonian $H$ (\ref{eq4}) 
as a function of eigenstate index  $j^\text{th}$. Here, the values
$\xi$, computed for eigenstates of both $\hat{n}$ and $\hat{\sigma}_z$,
are shown by black circles. Floquet states (ordered by increasing 
mean value of $\langle H\rangle$ averaged in a period) of (\ref{eq1}) 
are shown with red (gray) circles.
Here we have values $\omega_0=0.99$ ($\Delta_0=0.01$) and $\omega_0=0.975$ ($\Delta_0=0.025$) 
in top (a) and bottom (b) panels respectively with $g=0.04$, $\omega=1$, $\Omega=1.2$ 
and $f=\hbar\lambda\sqrt{n_p}$ with $\lambda=0.02$ and $n_p=20$. 
These values are the same as in panels (a) and (b) of Fig~\ref{fig4} respectively.
}
\label{fig7}
\end{figure}

The validity of the semiclassical description (\ref{eq7}) 
is confirmed by the numerical results presented in Fig.~\ref{fig6}
showing the dependence $h_0(h)$ for two spin (or qubit) projections.
Indeed, there is a good agreement between the numerical results 
obtained for the Hamiltonian (\ref{eq4}).

It is important to compare the numerical results obtained in the RWA of (\ref{eq4})
with the those  obtained from the Floquet eigenstates of (\ref{eq1}).
The index $j$ for Floquet eigenstates is defined for increasing value of 
$\langle H\rangle$ averaged over  a period.
We present the comparison for the participation ratio $\xi$ shown in Fig.~\ref{fig7}.
It shows a qualitative agreement between the Floquet results of (\ref{eq1})
and those obtained for the RWA Hamiltonian (\ref{eq4}). However,
the quantitative agreement is absent showing that $\xi$ values from RWA
are by a factor 2 different from Floquet values  of (\ref{eq1}).
We attribute this difference for the fact that the results are obtained in a close vicinity
to the resonance with $\omega_0 $ being very close to the driven frequency $\omega$.
In such a case next order corrections beyond RWA can produce additional frequency
shifts providing rescaling of an effecting value of frequency detuning
that would notably affect the values of participation ration $\xi$ of eigenstates.
We note that the difference between RWA an non-RWA cases in a resonance vicinity
was also pointed in \cite{refereeref2} even if the regime of strong oscillator
excitation was not analyzed in detail there.

\begin{figure}
\begin{center}
\includegraphics[width=0.46\textwidth]{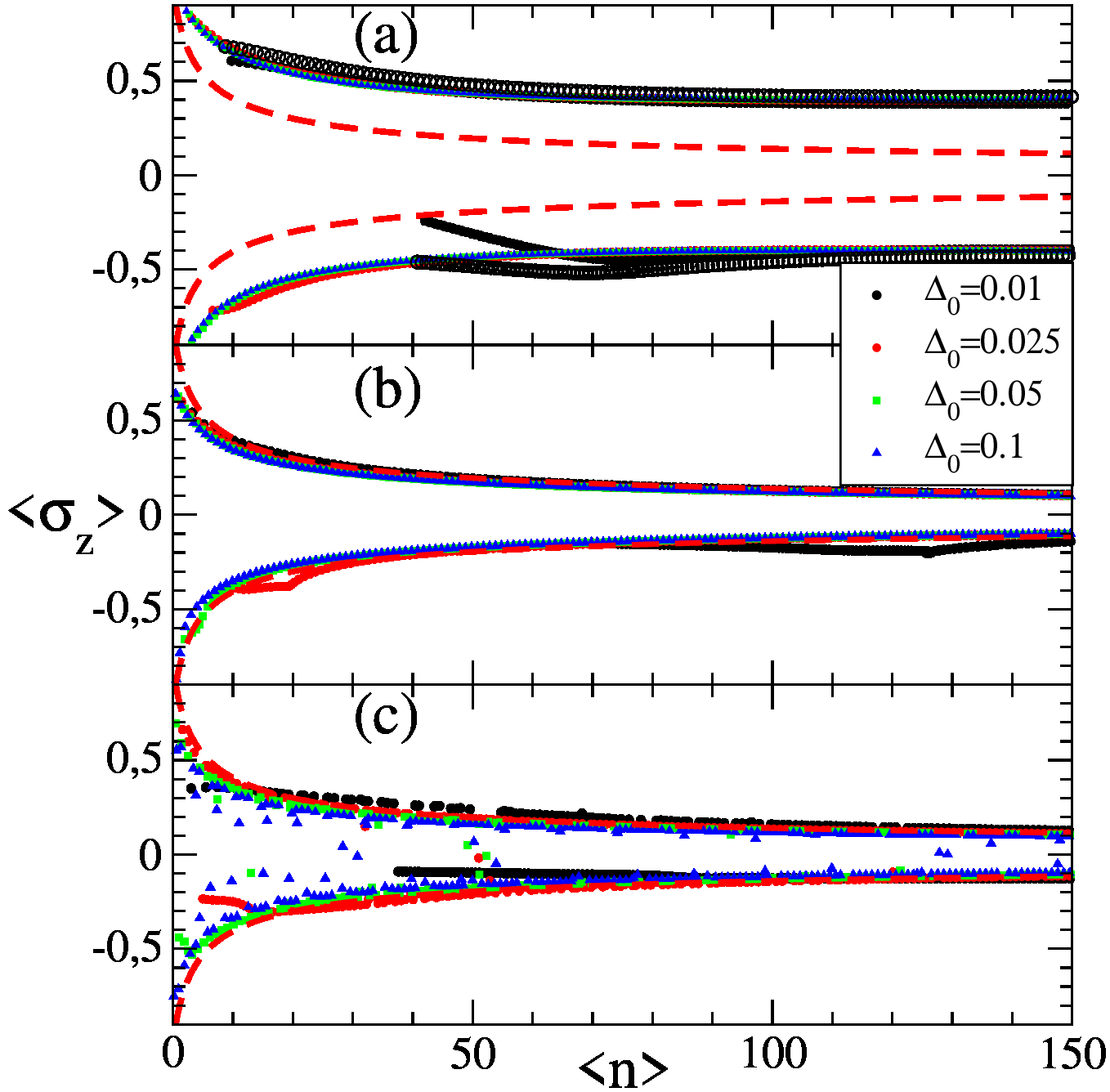}
\end{center}
\vglue -0.3cm
\caption{Average spin polarization as a function of mean oscillator 
number ($\langle\sigma_z\rangle$ vs. $\langle n\rangle$) for eigenstates of $H$.
Top (a), middle (b) and bottom (c) panels show the cases of Hamiltonian 
(\ref{eq1}), (\ref{eq4}) and \ref{eq5}) respectively.
Parameter values are $g=0.04$, $\omega=1$, $\Omega=1.2$ and $f=\hbar\lambda\sqrt{n_p}$ 
with $\lambda=0.02$ and $n_p=20$, with $\Delta_0=0.01, 0.025, 0.05, 0.1$ in 
black circles, red (gray) circles, green squares and blue triangles respectively.
The semiclassical theoretical dependence (\ref{eq8}) curve given by 
is shown by red dashed red (gray) curve for $\Delta_0=0.025$.
}
\label{fig8}
\end{figure}

According to the above argument the agreement between data obtained from  (\ref{eq1}),
(\ref{eq4}), (\ref{eq5}) should become better with the increase of resonance detuning
$\Delta_0$. We check this determining the dependence of average spin polarization 
$\langle\sigma_z\rangle$ on average quantum number of oscillator  $\langle n \rangle$
as it is presented in Fig.~\ref{fig8}. 
The comparison shows that the semiclassical theory (\ref{eq8})
well describes the numerical results of RWA from Hamiltonians of (\ref{eq4}),  (\ref{eq5}).
However, there is a notable deviations between the theory and RWA numerical results from
the Floquet results. At the same time, 
the results presented in Appendix Fig~\ref{figA3},~\ref{figA4}
show that the agreement between the Floquet results of (\ref{eq1})
and the RWA results of (\ref{eq4}) becomes better with in increase of 
resonance detuning $\Delta_0$ and decrease of coupling strength $g$.
This confirms our argument that the difference between
the Floquet  and RWA results are related to higher order corrections
related to coupling $g$ which play a more significant role in a close
vicinity to the resonance. 

\begin{figure}
\begin{center}
\includegraphics[width=0.46\textwidth]{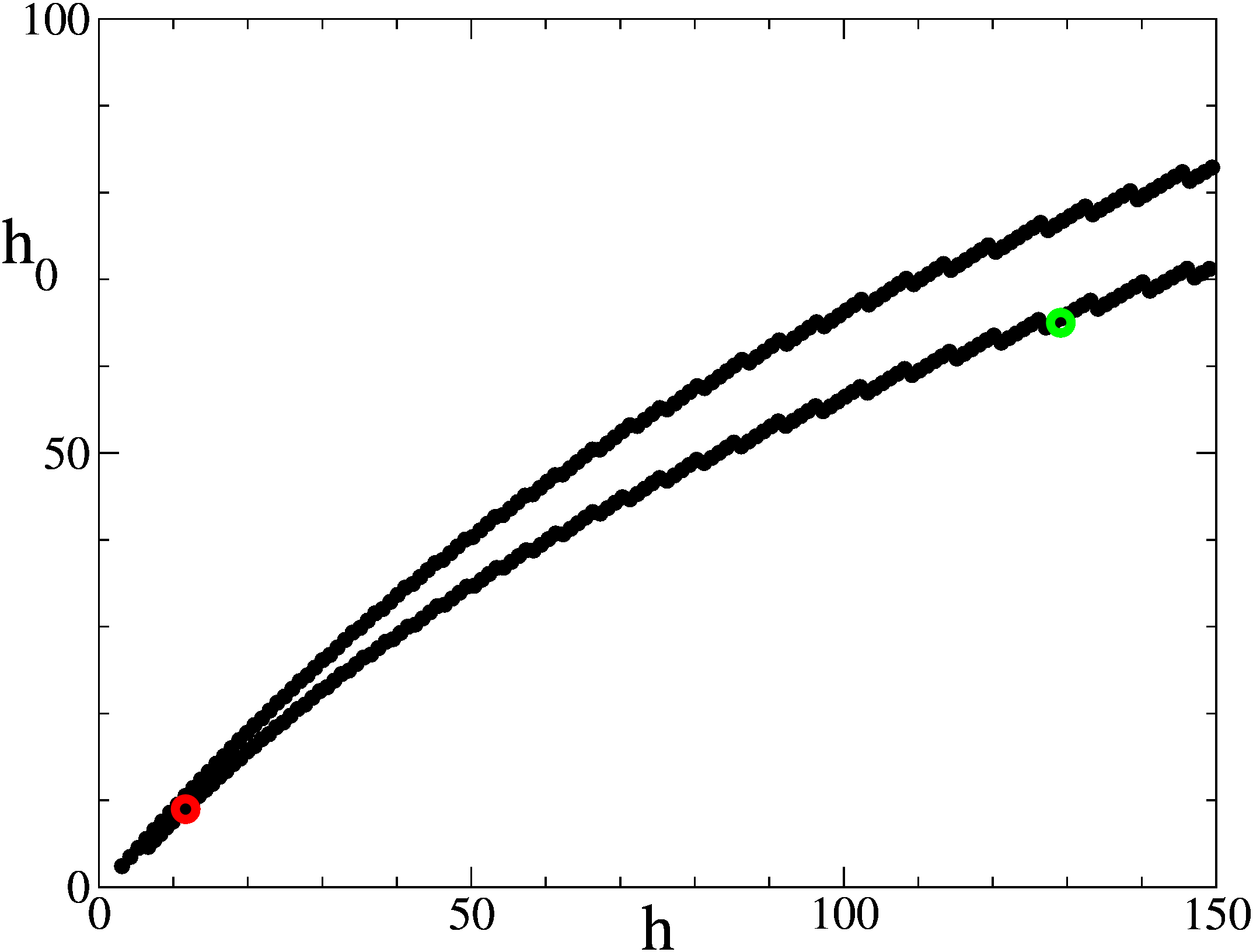}
\end{center}
\vglue -0.3cm
\caption{Quantum harmonic oscillator Hamiltonian 
$ h_0$ vs. $h$ for Floquet eigenstates $\vert\Psi_j(t=0)\rangle$. 
$h_0=\langle\Psi_j(t=0)\vert \hbar \omega \hat{n}\vert\Psi_j(t=0)\rangle$ 
and $h=\langle\Psi_j(t=0)\vert \hat{H}\vert\Psi_j(t=0)\rangle$ with 
$\hat{H}$ of Eq.\ref{eq1}
Parameter values are $g=0.04$, $\omega=1$, $\omega_0=0.975$, $\Omega=1.2$
and  $f=5^{-\frac{3}{2}}$ which are 
the same than panel $(b)$ of Fig~\ref{fig4}. Red and green circles represent 
the Floquet states shown in Fig~\ref{fig10} in top ($(a)$ and $(b)$) 
and bottom ($(c)$ and $(d)$) panels.
}
\label{fig9}
\end{figure}

In Fig.~\ref{fig9} we show the two branch dependence, 
corresponding to two spin polarizations,
of  quantities $h_0, h$ described above.
$h_0$ and $h$ of Fig.~\ref{fig9} are computed for Floquet 
eigenstates $\vert\Psi_j(t=0)\rangle$ valued in
initial state $t=0$ as 
$h_0=\langle\Psi_j(t=0)\vert \hbar \omega \hat{n}\vert\Psi_j(t=0)\rangle$ 
and $h=\langle\Psi_j(t=0)\vert \hat{H}\vert\Psi_j(t=0)\rangle$ 
where $\hat{H}$ is defined in Eq.\ref{eq1}.
We also mark with red and green circles there the 
values of $h_0, h$ obtained for two given Floquet 
states described in the next Section. 
The presence of two branches obtained from the developed semiclassical description
corresponds to the bistability behavior found in \cite{zsprl} for the evolution 
of Hamiltonian (\ref{eq1}) in presence of dissipation.

\section{Husimi function evolution} 
\label{sec5}

In this Section we consider the phase space representation in
the plane coordinate and momentum $(q,p)$ of 
certain Floquet eigenmodes of (\ref{eq1}) and 
the time evolution of certain initial coherent states.
The phase space representation of quantum states is done 
with the Husimi function which gives the Wigner function smoothed on 
a scale of Planck constant (see e.g.  \cite{husimi1,husimi2}).
The smoothing is done with the oscillator coherent state corresponding to a Gaussian
wave packet that is localized in the classical phase space around a point 
$(q_0,p_0)$ in the phase space. 
The smoothing is given by the relation
$\langle x|\varphi(p_0,q_0)\>=A e^{-(x-q_0)^2/2+\frac{i}{\hbar}p_0(x-q_0)}, $ 
with wave packet the same widths coordinate and momentum $\Delta p =\Delta q=1/2$, and with the normalization constant $A$
(see more details in \cite{husimi1,husimi2}).
Then the Husimi function probability $\rho_H$ in the phase space $(q_0,p_0)$ is given by
the relation $\rho_H(p_0,\theta_0)=|\<\varphi(p_0,\theta_0)\,|\,\psi\>|^2$.
We construct the Husimi function for up and down $\sigma_z$-spin components
of the total wavefunction.

\begin{figure}
\begin{center}
\includegraphics[width=0.46\textwidth]{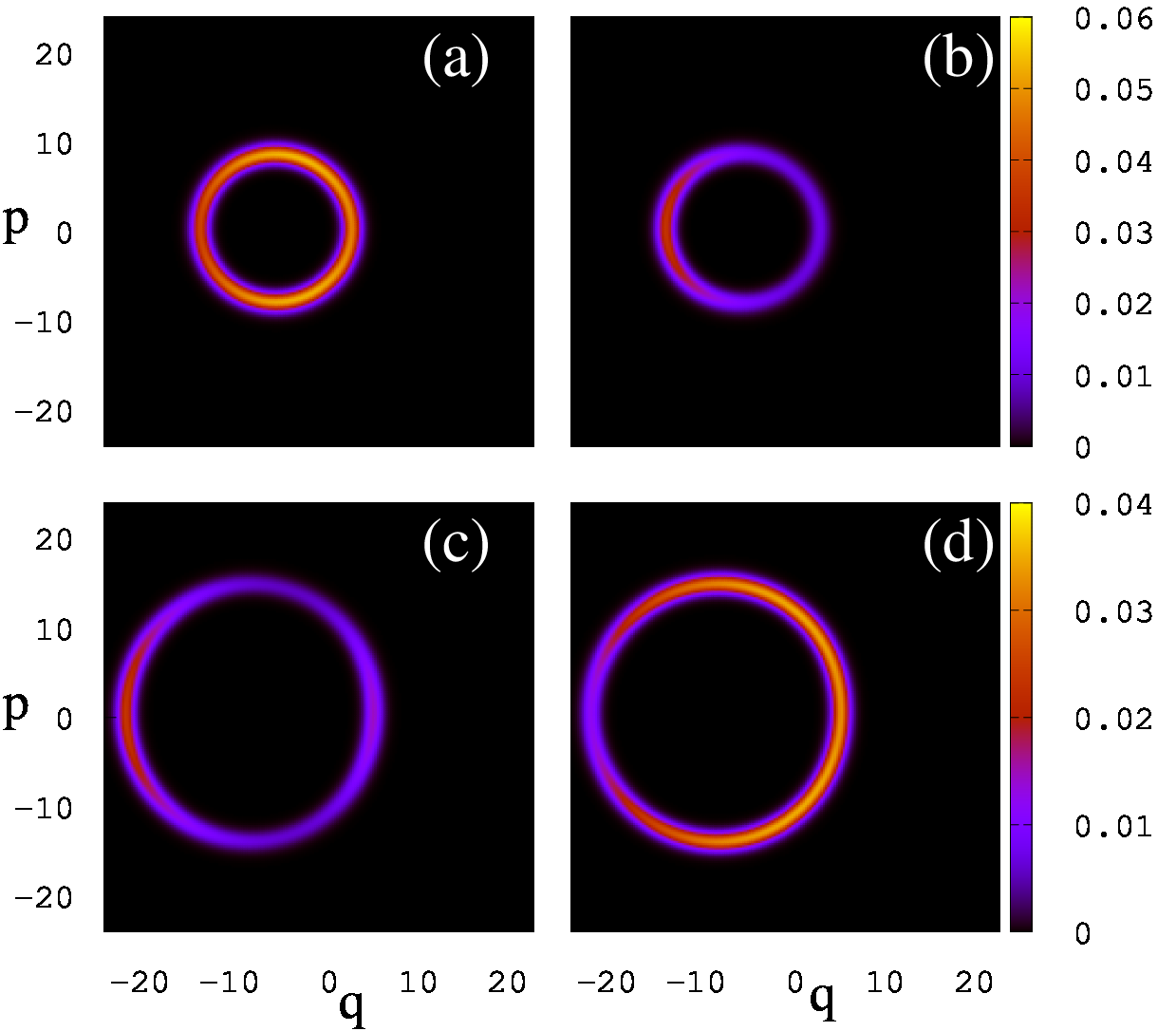}
\end{center}
\vglue -0.3cm
\caption{Husimi representation in phase space of two Floquet states with $t=0$. 
Left ($(a)$ and $(c)$) and right ($(b)$ and $(d)$) panels show 
the $\Pi_0\equiv\vert0\rangle\langle0\vert$ and $\Pi_1=\vert1\rangle\langle1\vert$
projection in $\hat{\sigma}_z$ respectively of two Floquet states.
The parameter values are the same as in Fig~\ref{fig9}, where both Floquet modes
are shown in ($h,h_0$) plane with color circles: 
red for $(a)$ and $(b)$ ($h\simeq11.64,h_0\simeq8.99$)
and green for $(c)$ and $(d)$ ($h\simeq129.13,h_0\simeq65.0$). 
The color map goes from black at 0
to yellow at maximum value given by 0.06 for $(a)$ and $(b)$, 
and 0.04 for $(c)$ and $(d)$.
}
\label{fig10}
\end{figure}

In Fig.~\ref{fig10} we present the Husimi functions for spin up and down
for a typical Floquet eigenstate with $\lambda=0.02$ and system parameters given in 
Fig.~\ref{fig9}. The results clearly show that the eigenstate
have double contribution of small and large oscillator numbers $n$
with a small circle in top panels and large circle in bottom panels
respectively (this doublet structure is present for both spin projections 
shown in left and right panels). This example shows that all phases 
of a circle in $(q,p)$ plane are present but the distribution over the phases
is inhomogeneous. The two sizes of the circle corresponds to
the two semiclassical branches appearing in (\ref{eq7}).

\begin{figure}
\begin{center}
\includegraphics[width=0.46\textwidth]{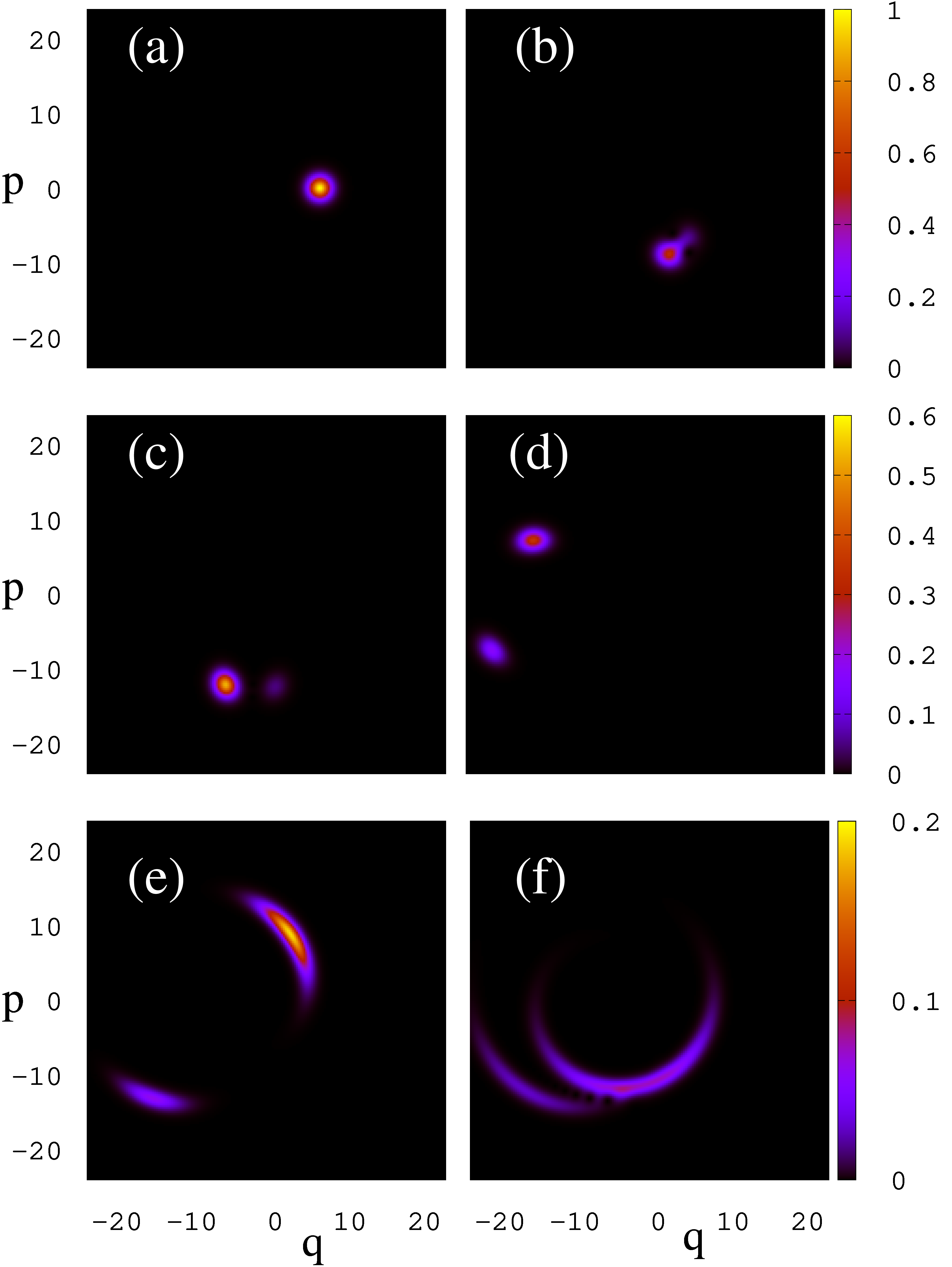}
\end{center}
\vglue -0.3cm
\caption{Husimi representation in phase space the evolution of a coherent state. 
Initial state is given by a coherent state centered at $(q_0,p_0)=(5,0)$ and spin projection 
$\Pi_0=\vert0\rangle\langle0\vert$ shown in panel (a). 
Time evolution of $\Pi_0$ projection is also shown for times
$\omega t /2\pi =10,20$, $50$ and $400$, $1000$ in $(b),(c)$, $(d)$ 
and $(e), (f)$ panels respectively.
Parameter values are $g=0.04$, $\omega=1$, $\omega_0=0.975$, $\Omega=1.2$
and $f=5^{-\frac{3}{2}}$ which are 
the same as in  panel $(b)$ of Fig~\ref{fig4}.
The videos of time evolution are available at \cite{suppmat}.
}
\label{fig11}
\end{figure}

The snapshots of time evolution of the Husimi function
of an initial coherent state are shown in Fig.~\ref{fig11}.
At large times the localized coherent state, 
shown in videos available at \cite{suppmat}, spreads over
the whole circle corresponding to a given oscillator number
that is in agreement with the Floquet eigenstate structure shown 
in Fig.~\ref{fig10} where the probability is distributed over
all circle phases even if the distribution
is inhomogeneous. 
The videos are obtained from the Floquet system (\ref{eq1}) and 
from the RWA Hamiltonian (\ref{eq2}).
The evolution in both cases is similar but not exactly the same.
More details about videos are given in Appendix.
The time of such a spreading $t_{sp}$ over the whole circle
is rather long with $\omega t_{sp}/2\pi \approx 1000$.

We attribute it to the nonlinear energy dispersion correction 
appearing in driven JCM due to coupling between the spin and oscillator
with $\delta \omega = \delta E_n \approx\pm  g \omega_0 \sqrt{n / \Delta_0} $
(see  (\ref{eq7})). In fact this nonlinear dependence of energy shift
$\delta E_n$ on level number $n$ (or classical action)
should lead to appearance of a nonlinear resonance 
with the driving frequency $\omega$. In principle, 
such a resonance can be treated in the pendulum approximation of an isolated 
resonance as described in \cite{chirikov}. 
Due to two spin orientations we will have  two resonances 
corresponding to spin up and down branches discussed above.
Thus there should exist a tunneling between this two branches with 
a certain tunneling time $\tau$.
The results presented in \cite{zsprl} (see Fig.5 there)
show that the tunneling times $\tau$, expressed in number of driving periods, 
can be rather long with  $\tau \sim 10^3 - 10^4$. We expect that the further development of
the nonlinear resonance theory
can allow to understand the mechanism of this long time tunneling process and obtain 
the estimates for its dependence on system parameters.
However, this requires to perform additional 
investigations going beyond the studies presented here. 
In the language of the Floquet eigenvalues the tunneling process should be related to 
appearing of very tiny splittings between Floquet eigenenergies $\epsilon_j$ in 
(\ref{eq3}). 

\section{Discussion} 
\label{sec6}

In this work we analyzed the JCM behavior under a monochromatic driving.
Our analytical and numerical results show that the system can be effectively reduced
to a modified JCM with a strong coupling between photons and qubit.
The obtained results allow to understand the process of two branches of excitation of 
many photons induced by the driving 
in presence of nonlinear frequency dispersion induced by
coupling between photons and qubit.
The obtained analytical semiclassical formula gives a good description
of obtained numerical results.
However, in a very close vicinity
of the resonance between frequencies of oscillator and
monochromatic driving there appear certain deviations
which we attribute to high order corrections to RWA
approach which become important in close
resonance vicinity. The obtained results still keep
certain open questions on properties on the driven JCM,
in particular the question about the physical estimates
of long tunneling times between two branches
corresponding to up and down qubit polarization,
which are also present in the dissipative case \cite{zsprl}.

Here we analyzed the case of unitary driven JCM system.
In experiments the dissipative effects start to play and important role.
However, at a weak dissipation the results obtained for the unitary evolution
will allow to have a better understanding of dissipative quantum behavior.
Thus our semiclassical theory for the unitary evolution
explains the appearance of bistability in the dissipative case  \cite{zsprl}.

Since the JCM is the fundamental system of quantum optics we hope that
the reach properties of driven JCM will attract interest
    of experimental groups working with superconducting qubits
    and other systems of quantum optics.

\begin{acknowledgments}
This work has been partially supported through the grant
NANOX $N^o$ ANR-17-EURE-0009 in the framework of 
the Programme Investissements d'Avenir (project MTDINA).
\end{acknowledgments}

\appendix*
\section{}
\label{appenda}

Here we present supplementary  
figures complementing the main text of the paper.  

\begin{figure}
\begin{center}
\includegraphics[width=0.48\textwidth]{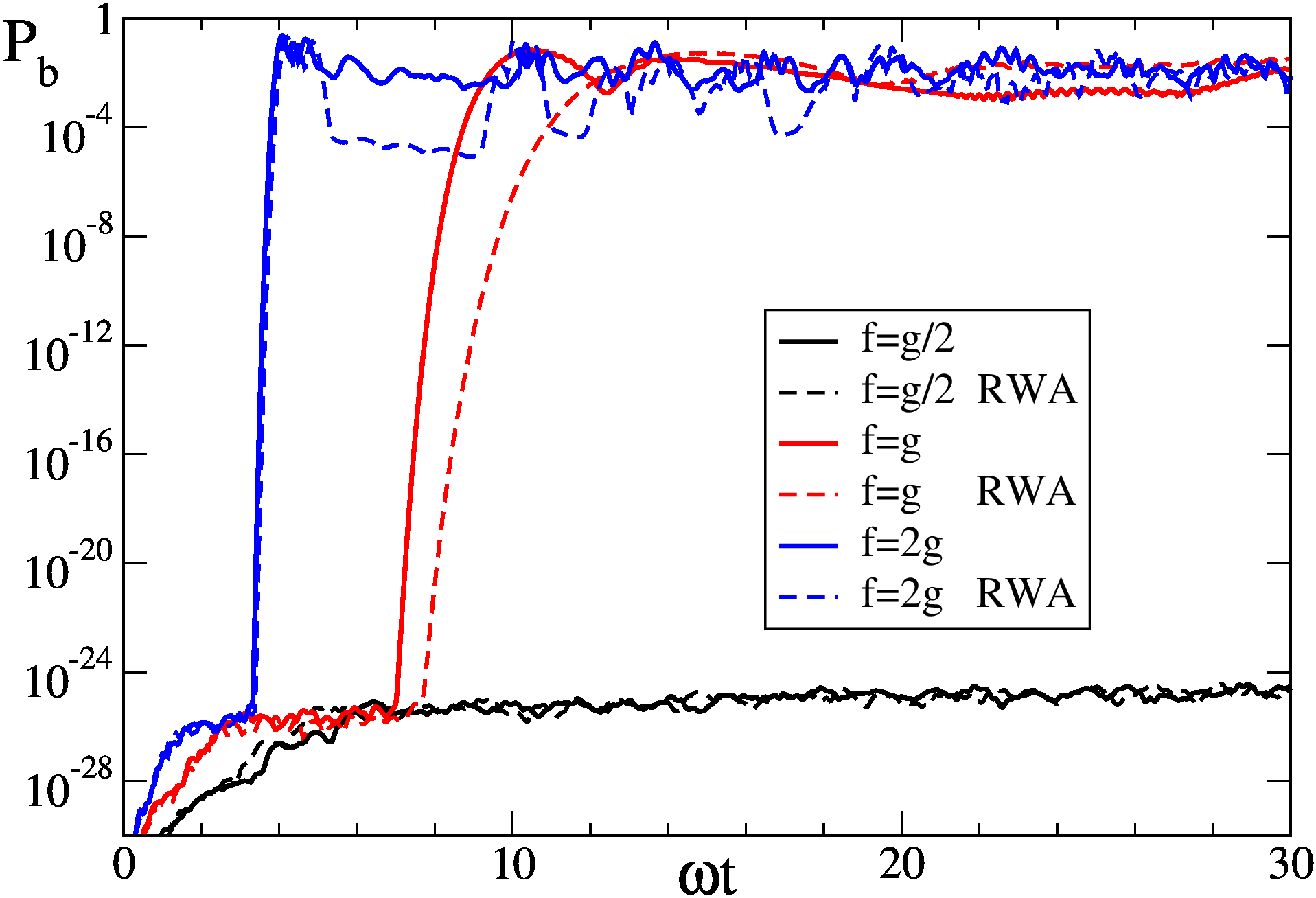}
\end{center}
\caption{\label{figA1}
Time evolution of total probability $P_b$ 
at 20 highest oscillator levels (with both spin components)
for numerical simulations
with basis of $1000$ oscillator states.
At the initial state the oscillator is at level $n=1$
and the spin is at its down-component.
Parameter values are $\omega=\omega_0=1$, $\Omega=1$, $g=0.04$ and $f=g/2$
(black curves), $f=g$ (red (gray) curves), $f=2g$ (blue (dark gray) curves),
where the dashed curves represent the computations within RWA frame
and full curves are for results without RWA for the Hamiltonian (\ref{eq1}).
Black curves are saturated at the level of $P_b \approx 10^{-24}$ at $\omega t =30$;
at the level $P_b \approx 10^{-16}$ the  blue (dark gray) curves
are located at $\omega t \approx 4$ and the red (gray) curves are  located
at $\omega t \approx 7.5$. 
}
\end{figure}

\begin{figure}
\begin{center}
\includegraphics[width=0.46\textwidth]{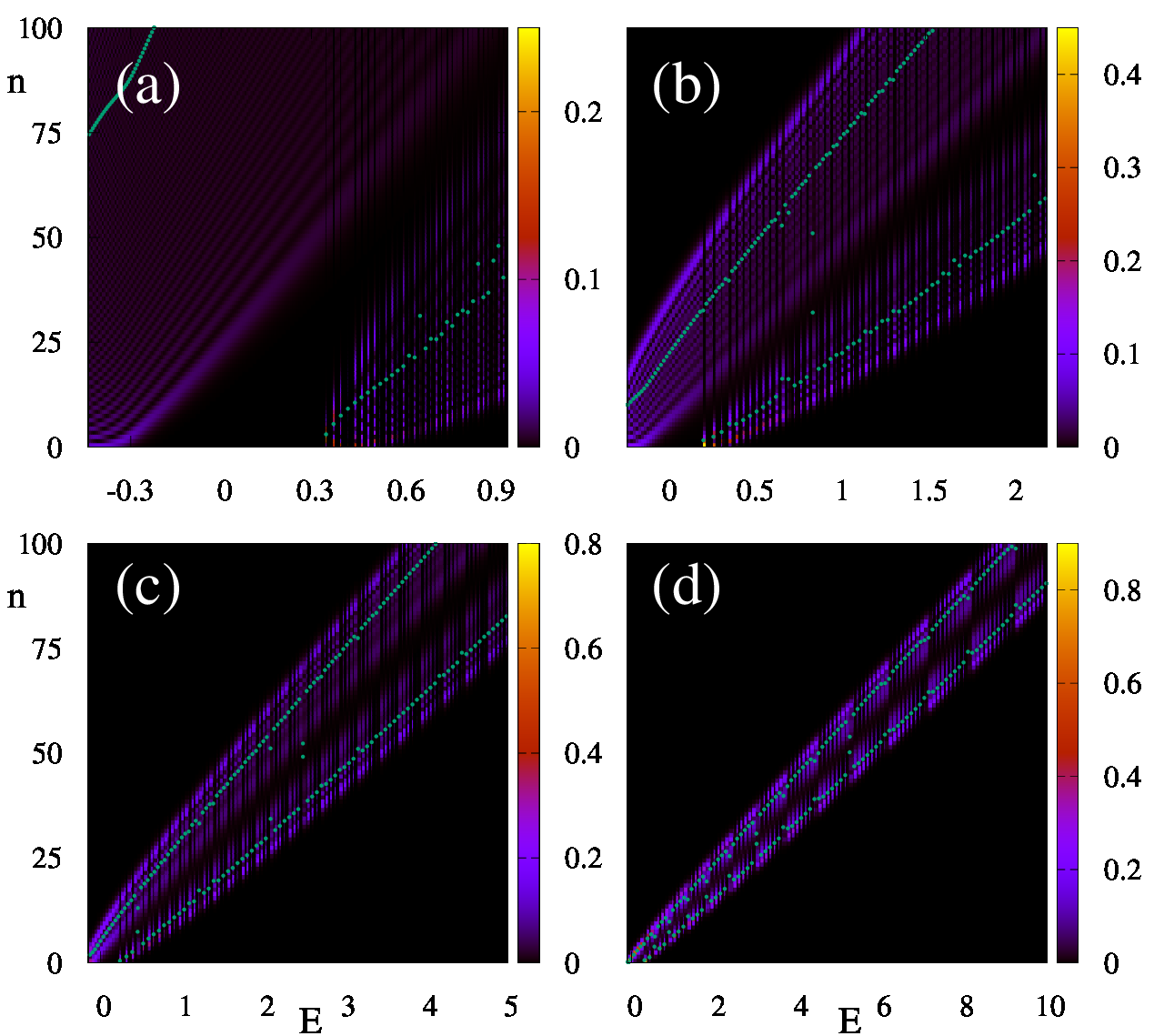}
\end{center}
\vglue -0.3cm
\caption{Same quantities and parameters as in Fig.~\ref{fig4}
but the eigenstates are obtained from the numerical diagonalization  of
transformed Hamiltonian (\ref{eq5}). The color map goes from black at 0
to yellow at maximum value given by $0.25$ for $(a)$, $0.45$ for $(b)$, 
$0.8$ for $(c)$ and $0.9$ for $(d)$.
}
\label{figA2}
\end{figure}

In Fig.~\ref{figA1} we show that for the case of exact resonance $\omega=\omega_0$
the probability is rapidly transferred to highest oscillator levels,
available for a given computational basis,
for $f \ge g$ while for $f < g$ the probability of high levels remains
very small. This numerical result is obtained both in RWA frame and without RWA 
for the Hamiltonian (\ref{eq1}).
Thus for $f \ge g$ 
the Floquet states are delocalized and non-nonrmalizable.
This result  is in agreement with the analytical  result obtained within RWA
in \cite{refereeref1}. Fig.~\ref{figA2} shows properties of eigenstates
of Hamiltonian (5) fog parameters of Fog.~\ref{fig4}.

\begin{figure}
\begin{center}
\includegraphics[width=0.48\textwidth]{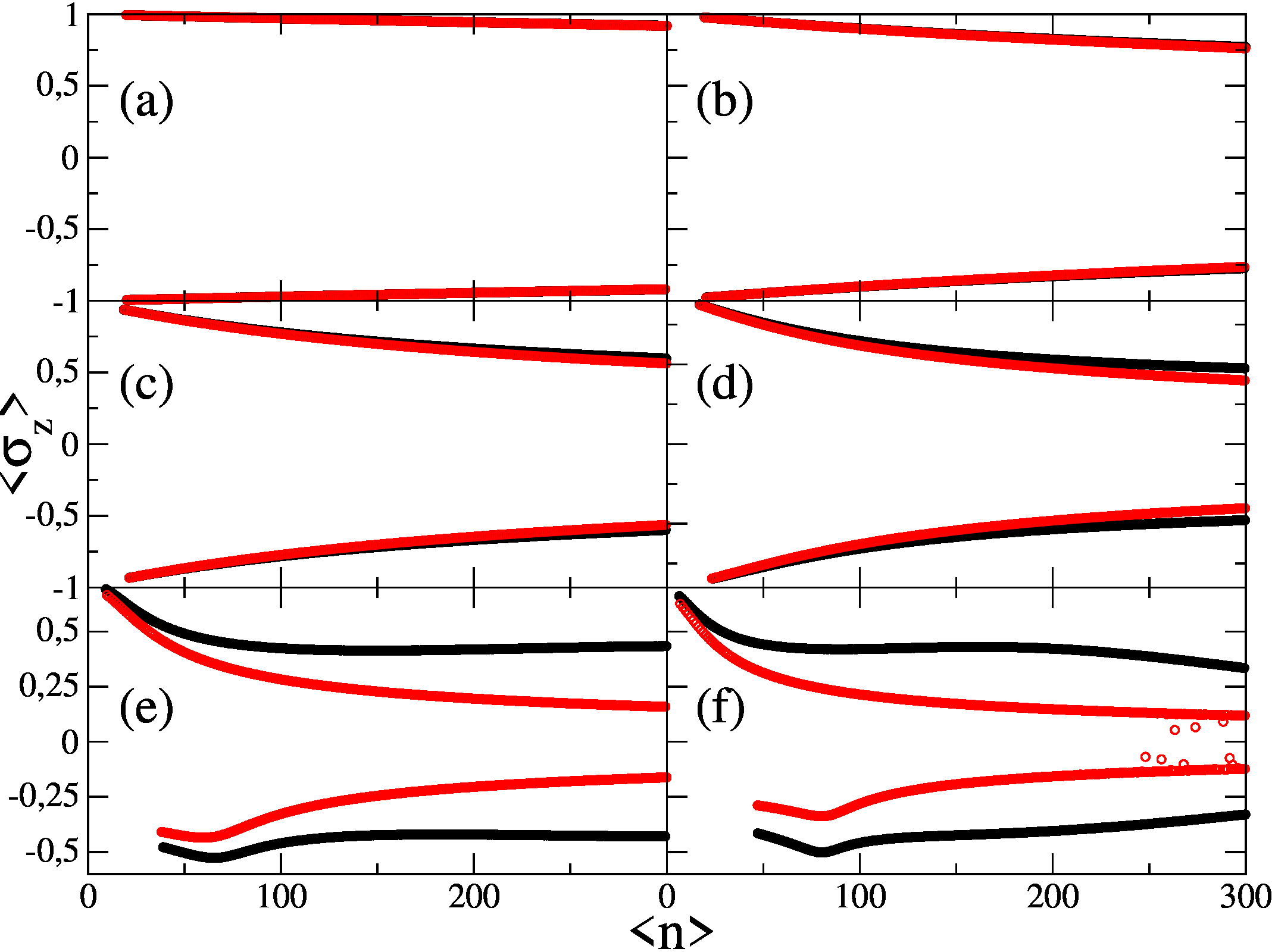}
\end{center}
\caption{\label{figA3}
$\langle\sigma_z\rangle$ vs. $\langle n\rangle$ for eigenstates of $H$
with $\Delta_0=0.01$ and different values of $g$.
Black and red (gray) circles show the cases of Hamiltonian 
of Eq.~\ref{eq1} and Eq.~\ref{eq4} respectively.
Parameter values are $\omega=1$, $\Omega=1.2$ and $f=5^{-\frac{3}{2}}$ with
a different value of $g$ en each panel: $0.0025$ $(a)$, 
0.005 $(b)$, 0.0088 $(c)$, 0.0138 $(d)$, 
0.0375 $(e)$ and 0.05 $(f)$.
}
\end{figure}

\begin{figure}
\begin{center}
\includegraphics[width=0.48\textwidth]{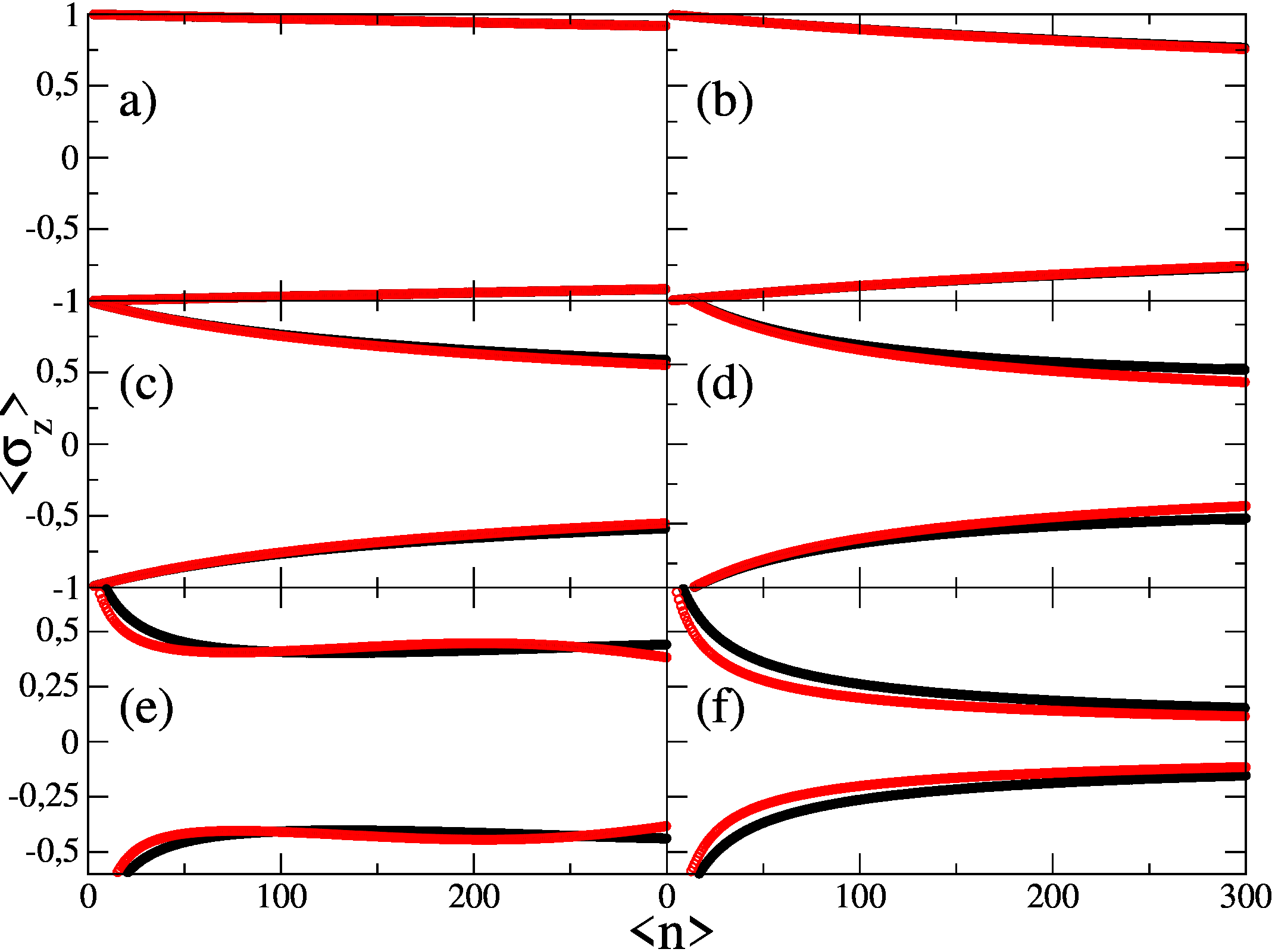}
\end{center}
\caption{\label{figA4} 
$\langle\sigma_z\rangle$ vs. $\langle n\rangle$ for eigenstates of $H$
with $\Delta_0=0.025$.
Black and red (gray) circles show the cases of Hamiltonian 
of Eq.~\ref{eq1} and Eq.~\ref{eq4} respectively.
Parameter values are the same as in Fig.~\ref{figA3} but with $\Delta_0=0.025$. 
Each panel represent a different value of $g=0.0025$ $(a)$, 
0.005 $(b)$, 0.0088 $(c)$, 0.0138 $(d)$, 
0.0375 $(e)$ and 0.05 $(f)$.
}
\end{figure}

Fig.~\ref{figA3} and Fig.~\ref{figA4} show the average spin polarization
as a function of the mean oscillator number 
($\langle\sigma_z\rangle$ vs. $\langle n\rangle$) for eigenstates of the Hamiltonian
of Eq.~\ref{eq1} (black circles) and Eq~\ref{eq4} (red circles)
with $\Delta_0=0.01$ and $\Delta_0=0.025$ respectively. 
Each panel on both figures represent a different value of $g$:
$0.0025$ $(a)$, $0.005$ $(b)$, $0.0088$ $(c)$, 
$0.0138$ $(d)$, $0.0375$ $(e)$ and $0.05$ $(f)$.

Videos in \cite{suppmat} present the time evolution of Husimi function
for parameters of Fig.~\ref{fig11}; videohusimi1.mp4
is obtained from the time evolution given by Floquet system (\ref{eq1})
and videohusimi2.mp4 is obtained from the RWA Hamiltonian (\ref{eq2}).
Initial state is given by a coherent state centered at $(q_0,p_0)=(5,0)$ 
with a spin projection in $\vert0\rangle$.
Parameter values are $g=0.04$, $\omega=1$, $\omega_0=0.975$, $\Omega=1.2$
and $f=\hbar\lambda\sqrt{n_p}$ with $\lambda=0.02$ and $n_p=20$.\\


\end{document}